\documentclass{llncs}
\usepackage{xcolor}
\usepackage{amssymb}
\usepackage{dsfont}
\usepackage{amsmath} 

\usepackage{url}
\usepackage{cancel}
\usepackage{verbatim}
\usepackage{alltt}
\usepackage{graphicx}

\usepackage{latexsym}
\usepackage{xspace}
\usepackage{comment}

\newcommand{\egdef}%
   {\ensuremath{\:\mathrel{\raisebox{-.7ex}%
   {$\stackrel{\rm def}{=\mkern-8mu=}$}}\:}}

\newcommand{\simlight}{\texttt{simlight}\xspace}

\newenvironment{ml}
  {\begin{alltt}
   \footnotesize} 
  {\end{alltt}
  }

\newenvironment{coq}
  {\begin{alltt}
   \footnotesize} 
  {\end{alltt}
  }

\newenvironment{humC}
  {\begin{alltt}
   \footnotesize}
  {\end{alltt}
  }

\begin{document}

\title{First Steps Towards the Certification \\ of an ARM Simulator
}
\titlerunning{Using Compcert-C for Certifying an ARM Simulator}
\authorrunning{X. Shi, F. Blanqui, J.-F. Monin \& F. Tuong}
\author{Xiaomu Shi\inst{1}
\and Fr\'ed\'eric Blanqui\inst{3}
\and Jean-Fran\c{c}ois Monin\inst{1,2}
\and Fr\'ed\'eric Tuong\inst{3}
}
\institute{%
Universit\'{e} de Grenoble 1 - LIAMA
\and CNRS - LIAMA
\and INRIA - LIAMA
}

\maketitle

\begin{abstract}
The simulation of Systems-on-Chip (SoC) is nowadays a hot topic
because, beyond providing many debugging facilities,
it allows the development of dedicated software before
the hardware is available.
Low-consumption CPUs such as ARM play a central role in SoC.
However, the effectiveness of simulation depends on the faithfulness
of the simulator. 
To this effect, we propose here to prove significant parts of
such a simulator, SimSoC.
Basically, on one hand, we develop a Coq formal model of the ARM 
architecture while on the other hand, we consider a
version of the simulator including components written in Compcert-C.
Then we prove that the simulation of ARM operations, 
according to Compcert-C formal semantics, 
conforms to the expected formal model of ARM.
Size issues are partly dealt with using automatic generation 
of significant parts of the Coq model and of SimSoC from
the official textual definition of ARM.
However, this is still a long-term project. 
We report here the current stage of our efforts
and discuss in particular the use of Compcert-C
in this framework.
\end{abstract}

\section{Introduction}
\label{sec:intro}

\subsection{Simulation of Systems-on-Chip}

Systems-on-Chip (SoC), used in devices such as smart-phones, contain both
hardware and software. A part of the software is generic and can be used with
any hardware systems, and thus can be developed on any computer. In contrast,
developing and testing the SoC-specific code can be done only with this SoC, or
with a \emph{software executable model} of the SoC. 
To reduce the time-to-market, the software development must start
before the hardware is ready. Even if the hardware is available,
simulating the software on a model provides more debugging
capabilities.

The 
fastest simulators use native simulation.
The software of the \emph{target} system (i.e., the SoC) is compiled with
the normal compiler of the computer running the simulator, 
but linked with special system
libraries. Examples of such simulators are the Android and iOS
SDKs.

In order to develop low-level system code, one needs a simulator that
can take the real binary code as input. Such a simulator requires a
model of the processor and of its peripherals (such as UART, DMA,
Ethernet card, etc). When simulating a smart-phone SoC, this kind of
functional simulator can be from 1 to 10 times slower than the real
chip.
These simulators have other uses, for example, as
reference models for the hardware verification.
An error in the simulator can then mislead both the software and the hardware
engineers.
QEMU~\cite{qemu} is an open-source processor emulator
coming with a set of device models; it can simulate several
operating systems. Other open-source simulators include
UNISIM~\cite{unisim} (accurately-timed)
and
SimSoC~\cite{ossc09}, developed by some colleagues, 
which is loosely-timed (thus faster).
Simics~\cite{simics} is a commercial alternative.
The usual language to develop such simulators is C++, combined with
the SystemC~\cite{systemc-lrm} and OSCI-TLM~\cite{tlm-osci}
libraries.

\medskip
The work reported here is related to SimSoC.

\subsection{The Need for Certification}

Altogether, a functional simulator is a complex piece of software.
SimSoC, which is able to simulate Linux both on ARM and PowerPC
architectures at a realistic speed
(over over 10 Millions instructions per second), 
includes about 60,000 lines of C++
code. The code uses complex features of the C++ language and of the
SystemC library. Moreover, achieving high simulation speeds requires
complex optimizations, such as dynamic translation~\cite{qemu}.

This complexity is problematic, because beyond speed,
\emph{accuracy} is required:
all instructions have to be simulated
exactly as described in the documentation.
There is a strong need to strengthen the confidence that
simulations results match the expected accuracy. 
Intensive tests are a first answer.
For instance, as SimSoC is able to run a Linux kernel
on top of a simulated ARM, we know that many situations are covered.
However it turned out, through further experiments, that it was not
sufficient: wrong behaviors coming from rare instructions
were observed after several months.
Here are the last bugs found and fixed by the SimSoC team while trying to boot
Linux on the SPEArPlus600 SoC simulator.
\begin{itemize}
\item After the execution of an \texttt{LDRBT} instruction, the contents of the
  base register (\texttt{Rn}) was wrong. It was due to a bug in the reference
  manual itself; the last line of the pseudo-code has to be deleted.
\item After a data abort exception, the base register write-back was not
  canceled.
\item Additionally, a half-word access to an odd address while
  executing some SPEArPlus600 specific code was not properly handled. 
\end{itemize}

Therefore we propose here to certify the simulator,
that is, to prove, using formal methods --
here: the Coq proof assistant \cite{coqmanual,coqart} --
that it conforms to the expected behavior.

This is a long term goal.
Before going to the most intricate features of a simulator such as SimSoC,
basic components have to be considered first.
We then decided to focus our efforts on a sensible and important 
component of the system: the CPU part of the ARMv6 architecture 
(used by the ARM11 processor family).
This corresponds to a specific component of the SimSoC simulator,
which was previously implementing the ARMv5 instruction set only.
Rather than certifying this component, it seemed to us more feasible
to design a new one directly in C, in such a way that it can be
executed alone, or integrated in SimSoC (by including the C code in
the existing C++ code).
We call this new component \simlight \cite{rapido11}. Combined with a small
\texttt{main} function, \simlight can simulate ARMv6 programs as long
as they do not access any peripherals (excepted the physical memory)
nor coprocessors. There is no MMU (Memory Management Unit) yet. Integrating it in SimSoC just
requires to replace the memory interface and to connect the interrupts
(IRQ and FIQ) signals.

\smallskip
The present paper reports our first efforts 
towards the certification of \simlight.
We currently have a formal description of the ARMv6 architecture,
a running version of \simlight,
and we are in the way of performing correctness proofs.
The standard way for doing this is to use Hoare logics
or a variant thereof.
Various tools exist in this area, 
for example Frama-C \cite{frama-c}.
We chose to try a more direct way,
based on an operational semantics of C;
more precisely the semantics of Compcert-C defined in the Compcert project
\cite{Leroy-Compcert-CACM}.
One reason
is that we look for a tight control on the formulation of proof obligations
that we will have to face.
Another advantage is that we can consider the use of
the certified compiler developed in Compcert,
and get a very strong guarantee on the execution of 
the simulator (but then, sacrificing speed to some extent\footnote{%
According to our first experiments, \simlight compiled with Compcert
is about 50~\% to 70~\% slower than \simlight compiled with \texttt{gcc -O0}.
}).

Another interesting feature of our work is that 
the most tedious (hence error prone) part of the formalization --
the specification of instructions --
is automatically derived from the reference manual.
It is well known that the formal specification of such big applications
is the main weak link in the whole chain.
Though our generators cannot be proved correct, because the statements
and languages used in the reference manual have no formal semantics,
we consider this approach as much more reliable than a manual formalization.
Indeed, a mistake in a generator will impact several or all operations,
hence the chances that it will be detected through a visibly wrong behavior
are much higher than with a manual translation,
where a mistake will impact only one (eventually rarely used) operation.

Note that after we could handle the full set of ARM instructions,
our colleagues of the SimSoC team decided to use the same technology
for the SimSoC itself:
the code for simulating instructions in \simlight,
i.e., the current component dedicated to the ARM v6 CPU in SimSoC,
is automatically derived using a variant of our generator,
whereas the previous version for ARM v5 was manually written \cite{rapido11}.

Fig.~\ref{fig:archi} describes the overall architecture. 
The contributions of the work presented in this paper are 
the formal specification of the ARMv6 instruction set
and the correctness proof of a significant operation.
More precise statements on the current achievements 
are given in the core of the paper.
 
\smallskip
\emph{Related Work.}
A fully manual formalization of the fm8501 and ARMv7 architectures are reported
in \cite{fm8501} and \cite{FoxM10}. 
The formal framework is respectively ACL2 and HOL4 instead of Coq, 
and the target is to prove that the hardware or microcode implementation of 
ARM operations are correct wrt the ARM specification.
Our work is at a different level:
we want to secure the simulation of programs using ARM operations.
Another major difference is the use of automatic generation 
from the ARM reference manual in our framework,
as stated above.

\medskip
The rest of the paper is organized as follows.
Section \ref{sec:overallarchi}
presents the overall architecture of \simlight
and indicates for which parts of \simlight 
formal correctness is currently studied.
A informal statement of our current results is also
provided there.
Sections \ref{sec:armmodel} and \ref{sec:simlight}
present respectively our Coq formal reference model of ARM
and the (Coq model of) Compcert-C programs targeted for correctness.
A precise statement of our current results and indications on the proofs are 
given in Section \ref{sec:results}.
We conclude in Section \ref{sec:conclusion} with some hints on
our future research directions.
Some familiarity with Coq is assumed in 
Sections \ref{sec:armmodel}, \ref{sec:simlight} and~\ref{sec:results}.


\newcommand{\SScert}{SimSoC-Cert\xspace}

\section{Main Lines of SimSoC-Cert}
\label{sec:overallarchi}

\subsection{Overall Architecture}
\begin{figure}
\hfil\includegraphics[width=.95\linewidth]{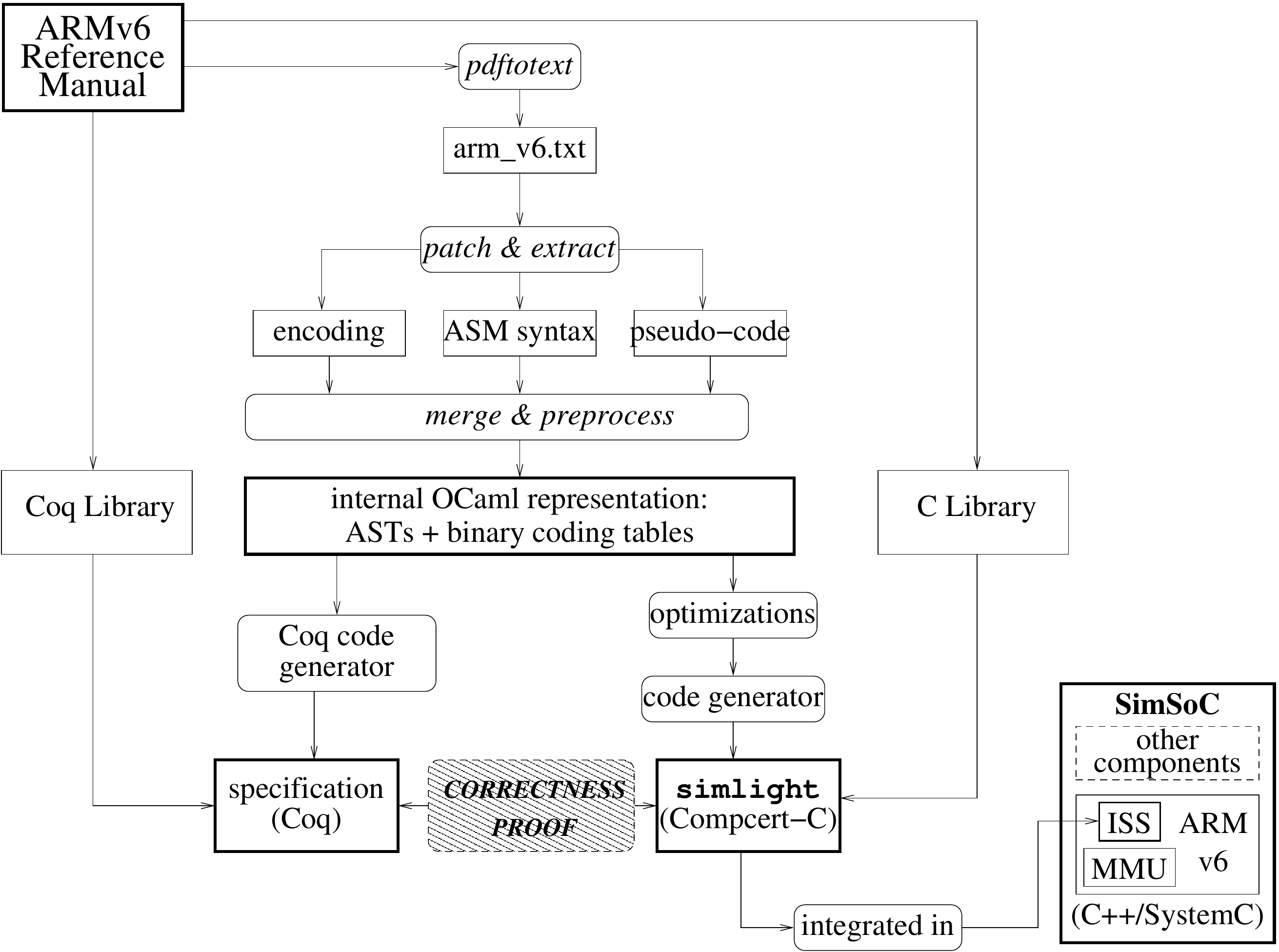}
\caption{Overall Architecture}
\label{fig:archi}
\end{figure}

The overall architecture of our system, called SimSoC-Cert, 
is given in Fig.~\ref{fig:archi}.
More specifically, we can see the data flow from
ARMv6 Reference Manual to the simulation code. 
Some patches are needed from the textual version of
the reference manual because the latter contains some minor bugs.
Three kinds of information are extracted for each ARM operation:
its binary encoding format, the corresponding assembly syntax
and its body, which is an algorithm operating on various
data structures representing the state of an ARM: registers, memory,
etc., according to the fields of the operation considered.
This algorithm may call general purpose functions defined
elsewhere in the manual,
for which we provide a Compcert-C library to be used by the simulator
and a Coq library defining their semantics.
The latter relies on Integers.v and Coqlib.v from CompCert library
which allows us, for instance, 
to manipulate 32-bits representations of words.
The result is a set of abstract syntax trees (AST) and binary
coding tables.
These ASTs follow the structure of the (not formally defined)
pseudo-code. 
Then two files are generated: 
a Coq file specifying the behavior of all operations
(using the aforementioned Coq library)
and a Compcert-C file to be linked with other
components of SimSoC (each instruction can also be executed
in standalone mode, for test purposes for instance).
More details are provided in \cite{rapido11}.

The decoding of ARM operations is not considered in the present paper:
this is important and planned for future work,
but is less urgent since we were already able 
to automatize the generation of intensive tests,
as reported in \cite{rapido11}. 
We therefore focus first on the algorithmic body of operations.
In order to state their correctness,
we need Coq ASTs for the Compcert-C statements of \simlight.
The code generator directly generates such ASTs.
Another option would be to work on a generated textual (ASCII)
presentation of the Compcert-C code,
but we prefer to avoid additional (and possibly unreliable)
parsing step as far as possible.
We will see in Section \ref{sec:simlight}
that these ASTs are moreover presented in a readable form using
suitable notations and auxiliary definitions.

The whole \simlight project is currently well-compiled by
Compcert (targeting Intel code) and gcc; moreover validation tests
succeed completely with both simulators.
The version of \simlight compiled with Compcert can serve 
as a reference simulator,
but for most purposes the version compiled with gcc is prefered
for its higher speed.



\subsection{Stating Correctness Theorems}

Let us now present the purpose of the gray box
of Fig.~\ref{fig:archi},
which represents our main target.

The correctness of simulated ARM operations is stated
with relation to the formal semantics of ARM
as defined by our Coq library and partly automatically produced
by the Coq code generator (the box called ``specification'' in
Fig.~\ref{fig:archi}).
Note that ARM operations are presented in a readable way using
suitable monadic constructs and notations:
apart from the security provided by automatic generation,
this greatly facilitates the comparison with the original pseudo-code
of the reference manual.
That said, it should be clear that the reference semantics
of ARM is the Coq code provided in these files.
Much effort has been spent in order to make them as clear
and simple as possible.

\begin{figure}
\hfil\includegraphics[width=.5\linewidth]{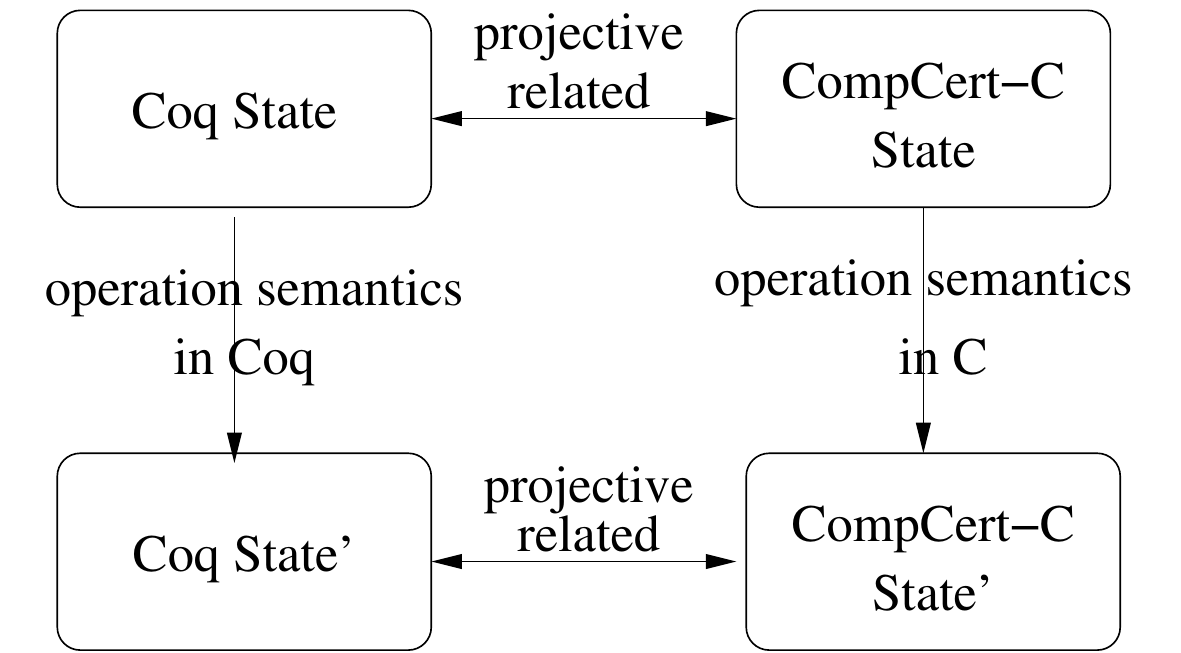}
\caption{Correctness of the simulation of an ARM operation}
\label{fig:thrm}
\end{figure}

In contrast, the Coq description of the behavior of corresponding
operations (as simulated by SimSoC -- Compcert-C programs)
is far more complicated, though the superficial structure
is quite similar. This will be detailed in Section~\ref{sec:simlight}.
In particular, the memory model of the latter version is much more 
complex.
In order to state correctness theorems, we define a relation
between 
an ARM state represented by the Compcert-C memory model
and 
another ARM state, as defined by the Coq formal semantics.
Essentially, we have a projection from the former to the latter.
Then for each ARM operation, we want a commutative diagram
schematized in Fig.~\ref{fig:thrm}.

For now, our automatic generation tools operate completely,
i.e., 
we have a Coq formal specification and a Compcert-C simulator 
for the full instruction set of ARM V6.
About proofs, 
the relationship between the abstract and the concrete memory models
is available;
we can then state correctness theorems for all ARM operations.
The work on correctness proofs themselves started recently.
We considered a significant ARM operation called \texttt{ADC}
(add with carry).
Our main theorem (Theorem 1 in Section \ref{sec:results})
states intuitevely that the diagram given
in Fig.~\ref{fig:thrm} commutes for \texttt{ADC}.
Its proof is completed up to some axioms on library functions,
details are given in Section~\ref{sec:results}.


\section{ARM Model}
\label{sec:armmodel}

\subsection{Processor Behavior}
A processor is essentially a transition system which operates
on a state composed of 
registers (including the program counter) and memory.
The semantics of its behavior amounts to repeat the following tasks:
fetch the binary code at a given address,
decode it as a processor operation and execute it;
the last task includes the computation of the address 
of the next operation.
The two main components of a processor simulator are then:
\begin{itemize}
\item The decoder, which, given a binary word, 
  retrieves the name of an operation and its potential arguments.
\item
  The precise description of transformations performed by
  an operation on registers and memory.
  In the reference manual of ARM, this is defined by
  an algorithm written in ``pseudo-code'' 
  which calls low-level primitives for, e.g., 
  setting a range of bits of a register to a given value.
  Some situations are forbidden or left unspecified.
  For  ARM processors, this results in
  a so-called ``UNPREDICTABLE'' state.
  The best choice, for a simulator, is then to stop with
  a clear indication of what happens.
\end{itemize}

Let us illustrate this on a concrete example.
Here is the original pseudo-code of the \texttt{ADC} (add with carry) operation of ARMv6.
As most operations of ARM, 
this operation has an argument called \texttt{cond} which
indicates whether the operation should be skipped or not.
\texttt{CPSR} (Current Program Status Register)
and 
\texttt{SPSR} (Saved Program Status Register, used for exception handling)
are special registers related to execution modes of ARM;
they also contain flags (\texttt{N}, \texttt{Z}, \texttt{C} and \texttt{V}) 
relevant to arithmetic instructions.
The instruction has four parameters:
\texttt{S} is a bit which specifies 
that the instruction updates \texttt{CPSR},
\texttt{Rn} is a register for the first operand,
\texttt{Rd} is the destination register,
\texttt{shifter\_operand} specifies the second operand
according to a (rather complicated) addressing mode.

\begin{humC}
A4.1.2 ADC
    if ConditionPassed(cond) then
        Rd = Rn + shifter_operand + C Flag;
        if S == 1 and d == 15 then
            if CurrentModeHasSPSR() then
                CPSR = SPSR;
            else UNPREDICTABLE
        else if S == 1 then
            N Flag = Rd[31];
            Z Flag = if Rd == 0 then 1 else 0;
            C Flag = CarryFrom(Rn + shifter_operand + C Flag);
            V Flag = OverflowFrom(Rn + shifter_operand + C Flag);
\end{humC}
\noindent
In the sequel, this version of \texttt{ADC} is referred to as
\texttt{ADC\_pseudocode}.
 

\subsection{Coq Semantics of ARM Operations}
\label{s:sem-ARM}

Each operation $O$ from the reference manual is mechanically translated 
to a corresponding Coq function named $O$\verb|_Coq|.
First we define a type 
\texttt{state}, which is a record with
two fields \texttt{Proc} and \texttt{SCC} (System Control Coprocessor)
containing respectively the components related to the main processor
(status register \texttt{CPSR}, \texttt{SPSR}, other registers...) 
and the corresponding components related to the coprocessor,
as well as the ARM memory model.
Then we use a monadic style~\cite{peyton-jones-tackling-09} in order to take 
the sequentiality of transformations on the state into account.
Beyond the state \texttt{st}, two other pieces of information are handled:
\texttt{loc}, which represent local variables of the operation
and \texttt{bo}, a Boolean indicating whether the program counter
should be incremented or not;
%
%
\noindent
they are registered in the following record
which is used for defining our monad:
\begin{coq}
Record semstate := mk_semstate
  \{ loc : local
  ; bo : bool
  ; st : state \}.

Inductive result \{A\} : Type :=
| Ok (_ : A) (_ : semstate)
| Ko (m : message)
| Todo (m : message).

Definition semfun A := semstate -> @result A.
\end{coq}
\noindent
Note that in general, every $O$\verb|_Coq| functions terminate with \texttt{Ok} as value. However for ``UNPREDICTABLE'' states for example, errors are implicitely propagated with our monadic constructors for exceptions : \texttt{Ko} and \texttt{Todo}.

We present now the translation of \texttt{ADC\_pseudocode}.
To this effect, we introduce \texttt{\_get\_st}, a monadic function
giving access to the current state \texttt{st} in its body, represented by
this notation:
\begin{coq}
Notation "'<' st '>' A" := (_get_st (fun st => A)) 
  (at level 200, A at level 100, st ident).
\end{coq}

\noindent
This yields the following code for \texttt{ADC\_Coq}:
\begin{coq}
(* A4.1.2 ADC *)
Definition ADC_Coq
   (S : bool) (cond : opcode)
   (d : regnum) (n : regnum) (shifter_operand : word) : semfun _ := <s0>
 if_then (ConditionPassed s0 cond)
   ([ <st> set_reg d (add (add (reg_content s0 n) shifter_operand)
                               ((cpsr st)[Cbit]))
   ;If (andb (zeq S 1) (zeq d 15))
     then    (<st> if_CurrentModeHasSPSR (fun em =>
       (<st> set_cpsr (spsr st em))))
     else         (if_then (zeq S 1)
       ([ <st> set_cpsr_bit Nbit ((reg_content st d)[n31])
       ; <st> set_cpsr_bit Zbit (if zeq (reg_content st d) 0 then repr 1 
                                 else repr 0)
       ; <st> set_cpsr_bit Cbit
                 (CarryFrom_add3 (reg_content s0 n) shifter_operand
                                 ((cpsr st)[Cbit]))
       ; <st> set_cpsr_bit Vbit (OverflowFrom_add3 (reg_content s0 n)
                                shifter_operand ((cpsr st)[Cbit])) ])) ]).
\end{coq}


\section{ARM Operations in Simlight}
\label{sec:simlight}

In the right branch of the overall architecture (Fig.~\ref{fig:archi}),
we generate \simlight according to the C syntax given by Compcert.
Here, actually we have two presentations of the corresponding code.
The first one is in a C source which is integrated into SimSoC 
(see \cite{rapido11} for more details), its contents is:
\begin{humC}
/* A4.1.2 ADC */
void ADC_simlight(struct SLv6_Processor *proc,
    const bool S,
    const SLv6_Condition cond,
    const uint8_t d,
    const uint8_t n,
    const uint32_t shifter_operand)
\{
const uint32_t old_Rn = reg(proc,n);
if (ConditionPassed(&proc->cpsr, cond)) \{
 set_reg_or_pc(proc,d,((old_Rn + shifter_operand) + proc->cpsr.C_flag));
 if (((S == 1) && (d == 15))) \{
  if (CurrentModeHasSPSR(proc))
   copy_StatusRegister(&proc->cpsr, spsr(proc));
  else
   unpredictable();
  \} else \{
   if ((S == 1)) \{
    proc->cpsr.N_flag = get_bit(reg(proc,d),31);
    proc->cpsr.Z_flag = ((reg(proc,d) == 0)? 1: 0);
    proc->cpsr.C_flag = 
     CarryFrom_add3(old_Rn, shifter_operand, proc->cpsr.C_flag);
    proc->cpsr.V_flag = 
     OverflowFrom_add3(old_Rn, shifter_operand, proc->cpsr.C_flag);
\} \} \} \}
\end{humC}
This piece of code uses a function called \texttt{set\_reg\_or\_pc}
instead of \texttt{set\_reg}:
the latter also exists in \simlight and the function to be used depends
on tricky considerations about register 15, which happens to be the PC.
More details about this are given in Section~\ref{s:diff-sem-simlight}.

The second presentation is an AST according to a Coq inductive type defined
in Compcert.
\begin{coq}
Definition ADC\_Coq\_simlight := (ADC, Internal 
    \{| fn_return := void; fn_params := 
      [proc -: `*` typ_SLv6_Processor; 
       S -: uint8; cond -: int32; d -: uint8; n -: uint8; 
       shifter_operand -: uint32];
      fn_vars := [ old_Rn -: uint32];
      fn_body :=
($ old_Rn`:\begin{math}\circ\end{math}) `= (call (\textbackslash{}reg`:\begin{math}\circ\end{math}) E[\textbackslash{}proc`:\begin{math}\circ\end{math}; \textbackslash{}n`:\begin{math}\circ\end{math}] \begin{math}\circ\end{math})`:\begin{math}\circ\end{math};;
`if (\begin{math}\bullet\end{math} (\textbackslash{}ConditionPassed`:\begin{math}\circ\end{math}) E[&((`*(\textbackslash{}proc`:\begin{math}\circ\end{math})`:\begin{math}\circ\end{math})|cpsr`:\begin{math}\circ\end{math})`:\begin{math}\circ\end{math}; \textbackslash{}cond`:\begin{math}\circ\end{math}] \begin{math}\circ\end{math})
 then (\begin{math}\bullet\end{math} (\textbackslash{}set_reg_or_pc`:\begin{math}\circ\end{math}) E[\textbackslash{}proc`:\begin{math}\circ\end{math}; \textbackslash{}d`:\begin{math}\circ\end{math}; ((\textbackslash{}old_Rn`:\begin{math}\circ\end{math})+\begin{math}\bullet\end{math})+\begin{math}\bullet\end{math}:\begin{math}\circ\end{math}] \begin{math}\circ\end{math});;
 `if ((($ S`:\begin{math}\circ\end{math})==(#1`:\begin{math}\circ\end{math})`:\begin{math}\circ\end{math})&(($ d`:\begin{math}\circ\end{math})==(#15`:\begin{math}\circ\end{math})`:\begin{math}\circ\end{math})`:\begin{math}\circ\end{math})
  then `if (call (\textbackslash{}CurrentModeHasSPSR`:\begin{math}\circ\end{math}) E[\textbackslash{}proc`:\begin{math}\circ\end{math}] \begin{math}\circ\end{math})
   then (call (\textbackslash{}copy_StatusRegister`:\begin{math}\circ\end{math}) E[&(\begin{math}\bullet\end{math}|cpsr`:\begin{math}\circ\end{math})`:\begin{math}\circ\end{math}; \begin{math}\bullet\end{math}] \begin{math}\circ\end{math})
   else (call ($ unpredictable`:\begin{math}\circ\end{math}) E[] \begin{math}\circ\end{math})
  else `if (($ S`:\begin{math}\circ\end{math})==(#1`:\begin{math}\circ\end{math})`:\begin{math}\circ\end{math})
   then ((($ proc`:\begin{math}\circ\end{math})|cpsr`:\begin{math}\circ\end{math})|N_flag`:\begin{math}\circ\end{math}) `= 
    (\begin{math}\bullet\end{math} (\textbackslash{}get_bit`:\begin{math}\circ\end{math}) E[(\begin{math}\bullet\end{math} (\textbackslash{}reg`:\begin{math}\circ\end{math}) E[\textbackslash{}proc`:\begin{math}\circ\end{math}; \textbackslash{}d`:\begin{math}\circ\end{math}] \begin{math}\circ\end{math}); #31`:\begin{math}\circ\end{math}] \begin{math}\circ\end{math})`:\begin{math}\circ\end{math};;
        ((($ proc`:\begin{math}\circ\end{math})|cpsr`:\begin{math}\circ\end{math})|Z_flag`:\begin{math}\circ\end{math}) `= 
    (((\begin{math}\bullet\end{math} (\textbackslash{}reg`:\begin{math}\circ\end{math}) \begin{math}\bullet\end{math} \begin{math}\circ\end{math})==(#0`:\begin{math}\circ\end{math})`:\begin{math}\circ\end{math})?(#1`:\begin{math}\circ\end{math})`:(#0`:\begin{math}\circ\end{math})`:\begin{math}\circ\end{math})`:\begin{math}\circ\end{math};;
        ((($ proc`:\begin{math}\circ\end{math})|cpsr`:\begin{math}\circ\end{math})|C_flag`:\begin{math}\circ\end{math}) `= 
    (\begin{math}\bullet\end{math} (\textbackslash{}CarryFrom_add3`:\begin{math}\circ\end{math}) E[\begin{math}\bullet\end{math}; \begin{math}\bullet\end{math}; (\begin{math}\bullet\end{math} (\begin{math}\bullet\end{math}|C_flag`:\begin{math}\circ\end{math}) \begin{math}\circ\end{math})] \begin{math}\circ\end{math})`:\begin{math}\circ\end{math};;
        ((($ proc`:\begin{math}\circ\end{math})|cpsr`:\begin{math}\circ\end{math})|V_flag`:\begin{math}\circ\end{math}) `= 
    (\begin{math}\bullet\end{math} (\textbackslash{}OverflowFrom_add3`:\begin{math}\circ\end{math}) E[(\begin{math}\bullet\end{math} (\textbackslash{}old_Rn`:\begin{math}\circ\end{math}) \begin{math}\circ\end{math}); \begin{math}\bullet\end{math}; \begin{math}\bullet\end{math}] \begin{math}\circ\end{math})`:\begin{math}\circ\end{math}
   else skip
 else skip |\}).
\end{coq}
%
The symbols ``$\circ$'' and ``$\bullet$'' do not belong to
the real notations, they stand for types and sub-terms not represented here
for the sake of simplicity.
Indeed, an important practical issue is that
Compcert-C ASTs include types everywhere,
hence a naive approach would generates heavy and repetitive
expressions at the places where $\circ$ occurs,
thus making the result unreadable
(and space consuming).
We therefore introduce auxiliary definitions for types
and various optimizations for sharing type expressions.
We also introduce additional convenient notations,
as shown above for \texttt{ADC\_Coq\_simlight},
providing altogether a C-looking presentation of the AST.

We plan to generate the first form from the AST using a pretty-printer. 
The following discussion is based on the AST presentation.

\subsection{Differences with the Coq Model of ARM Operations}
\label{s:diff-sem-simlight}

Although the encoding of operations in \simlight and 
in the Coq semantics of ARM are
generated from the same pseudo-code AST, results are rather
different because, on one hand, they are based on different data types
and, on the other hand, their semantics operates on different memory models.
Therefore, the proof that the simulation of an operation in \simlight 
behaves as expected according to the Coq semantics is not trivial.

In the Coq model of ARM, everything is kept as simple as possible. 
ARM Registers are presented by words, 
the memory is a map from address to contents, 
the initial value of parameters such as \texttt{Rn} 
is available for free -- we are in an functional setting,
etc.
In contrast, \simlight uses an imperative setting
(hence the need to store the initial value of  \texttt{Rn}
in  \texttt{Old\_Rn}, for instance).
More importantly, complex and redundant data structure are involved
in order to get fast speed.
For example, a 32 bits wide status register is defined as a data structure
containing, for every significant bit, a field of Boolean type
-- internally, this is represented by a byte. 
A more interesting example is the program counter, which is
the register 15 at the same time.
As this register is sometimes used as an ordinary register,
and sometimes used as the PC, 
the corresponding data structure implemented in \simlight
includes an array which stores all the registers
and a special field \texttt{pc}, which is a pointer aliasing register 15.
This register plays an important role in ARM architecture.
Its value is used in the \texttt{may\_branch} condition 
for simulating basic blocks~\cite{rapido11}. 
And during the simulation loop, it is read many times.
Note that this special field \texttt{pc} is read-only.

Moreover we have to work with the Compcert memory model of such
data structures. This, model detailed in \cite{lerbla08}, 
introduces unavoidable complications
in order to take low-level considerations,
such as overlapping memory blocks into account.
Another source of complexity is that, in a function call,
a local variable takes a meaningful value only after
a number of steps representing parameter binding.
More details are given in Section~\ref{sec:results}.

Another important difference is that,
in the Coq specification, the semantics is defined by a function
wheras, 
in Compcert-C, the semantics is a relation between the
initial memory and the final memory when evaluating statements
or expressions.

\subsection{Translation from Pseudo-code AST to Compcert-C AST}

We describe here the mapping from the pseudo-code AST to Compcert-C AST.
This translation is not only to Compcert-C AST, 
but more specifically to the Compcert-C AST for \simlight. 
It makes use of an existing library of functions dedicated to \simlight.
For example in \texttt{ADC\_pseudocode}, 
the occurrence of \texttt{CPSR}
stands for an expression representing 
the contents of \texttt{CPSR} in the current state.
But in \simlight, 
this corresponds to a call to a library function 
\texttt{StatusRegister\_to\_uint32}.
The translation deals with many similar situations.


Let us now sketch the translation process.
Both the definitions of pseudo-code AST and Compcert-C AST
include inductive types for expressions, statements and programs.
Compcert-C expressions are limited to common programming operations 
like binary arithmetic operations, type cast,
assignments, function calls, etc. 
For many constructors of pseudo-code AST the mapping is quite natural,
but others require a special treatment:
the ones which are specific to ARM, 
for representing registers, memory and coprocessor expressions,
invocation of library functions, or bit range expressions.
Those special expressions are translated to Compcert-C function calls.
For example, the pseudo-code expression \texttt{Reg (Var n, Some m)},
designates the contents of register number \texttt{n} with the ARM processor mode \texttt{m}.
In \simlight, this becomes a call to \texttt{reg\_m} with parameters \texttt{proc}, \texttt{n} and \texttt{m}.

In summary, the translation of expressions looks as follows:
\begin{ml}
let rec Transf\_exp = function
  | Reg (e, m) -> Ecall reg\_m ...
  | CPSR -> Ecall StatusRegister\_to\_uint32 ...
  | Memory (e, n) -> Ecall read\_mem ...
  | If\_exp (e1, e2, e3) -> Econdition ...
  | BinOp (e1, op, e2) -> Ebinop ...
  | Fun (f, e) -> Ecall f ...
  ...
\end{ml}

For statements, we have a similar situation.
Here, assignments require a special attention.
For example in \texttt{ADC\_pseudocode}, there is an assignment
\texttt{CPSR=SPSR}. 
In \simlight, this assignment is dealt with 
using a call to the function \\\texttt{copy\_StatusRegister}.
The corresponding Compcert-C AST embeds this call
as an argument of the constructor \texttt{Sdo}.

In summary, the translation of statements looks as follows:
\begin{ml}
let rec Transf\_stm = function
  | Assign (dst, src) -> Sdo (Ecall funct ...)
  | For (c, min, max, i) -> Sfor ...
  | If (e, i1, i2) -> Sifthenelse ...
  | Case (e, s, default) -> Sswitch ...
  ...
\end{ml}

In our case,
each operation is transformed to a Compcert-C program
where there are no global variables,
the function list contains only 
the function corresponding to the considered ARM operation
(let us call it $f$, it is of course an internal function), 
and with an empty \texttt{main}.
When the program is called, 
the global environment built for this program will
only contain a pointer to $f$.
 
The translation from pseudo-code AST program to Compcert-C AST program
has the following shape:
\begin{ml}
let Transformed\_program = 
  \{ vars = []
  ; functs = [ Internal (instr\_id,
                  \{ fn\_return = Tvoid
                  ; fn\_params = ... (* operation parameters *)
                  ; fn\_vars = ... (* operation local variables *)
                  ; fn\_body = ... (Transf\_stm ...) \}) ]
  ; main = empty\_main \}
\end{ml}


\section{Current Proofs}
\label{sec:results}
\newcommand{\ofs}{\ensuremath{\mathit{ofs}}}

On both sides, the Compcert-C \simlight model and the Coq ARM model, 
the state of the processor is expressed by a big Coq term.
In the Compcert-C \simlight model, the processor state information is gathered in
a data structure \texttt{SLv6\_Processor}, 
which includes the MMU, 
the status registers \texttt{CPSR} and \texttt{SPSR}, the
system coprocessor and the registers.
In the Coq formal model of ARM, the processor state is represented
by a value of type 
\texttt{result}, described in Section~\ref{s:sem-ARM}.

It is clearly possible to define a projection from
a \texttt{SLv6\_Processor} $M$
to a \texttt{result} $r$.
Then we say that $M$ and $r$ are \emph{projective-related},
denoted by $\mathit{proc\_state\_related}\;M\;r$.
The evil is in the details of the different type definitions,
especially for the memory models.
Here are the guiding ideas.
Once a function such as \texttt{ADC\_Coq\_simlight} is called, 
parameters are allocated in memory, and a local environment is built. 
This local environment contains the mapping
from identifiers to a memory block reference.
For a variable of type \texttt{struct}, such as the ARM processor, 
the environment only yields an entry pointer to the structure.
Here, the type information generated for our Compcert-C AST
is needed in order to find fields inside Compcert-C memory,
and to retrieve the processor model.
The main function used there from Compcert is \verb|load|. 
Its arguments are a memory $M$, a block $b$, 
an offset $\ofs$ and the type $\tau$ of the value to be loaded
from $b$ at $\ofs$. 
Other variables who have a simple type like \texttt{int32},
are directly accessed by their identifier from the environment.


Let us now consider a specific instance of Fig.~\ref{fig:thrm},
applied to \texttt{ADC}.
We choose it first because it is a typical ARM operation, 
which involves various ways of changing the processor state,
and arithmetic calculations.
Moreover, all data-processing operations have a very similar structure. 
If we prove the correctness of the \simlight implementation of \texttt{ADC}, 
we can expect to automate the proofs for the others data-processing operations.

The proof exploits the formal operational semantics of Compcert-C,
which is defined as a transition system
$$G,E\vdash \langle\textrm{piece of code}\rangle,M\overset{t}{\Rightarrow}out,M'$$
where $G$ represents the global environment (constants) of the program,
$E$ represents the local environment,
$M$ and $M'$ represent memory states,
$t$ is the trace of input/output events
and $out$ is the outcome.
In our case, the piece of code is \texttt{ADC\_Coq\_simlight},
and the trace of input/output event ($t$) is empty:
all function calls are internal calls.
Compcert-C offers two kinds of operational semantics:
small-step and big-step semantics.
The latter is better suited to our needs because 
the statement of correctness, along the diagram in Fig.~\ref{fig:thrm},
relates states before and after the execution of the body of an operation.
The precise statement of our theorem is as follows.
\begin{theorem}\label{t:adc}
Let $M$ and $M'$ be the memory contents respectively before and
after the simulation of \texttt{ADC\_Coq\_simlight};
similarly, let $st$ and $st'$ be the state of ARM in its formal model.
If $M$ and $st$ are projective-related,
as well as the arguments of the call to \texttt{ADC},
then $M'$ and $st'$ are projective-related as well.
Formally, if:
\begin{itemize}
\item $\mathit{proc\_state\_related} \: M \: (\texttt{Ok}\: st)$
\item similarly for the arguments of \texttt{ADC}
\item 
  $G,E\vdash \texttt{ADC\_Coq\_simlight},M\overset{t}{\Rightarrow}out,M'$
\end{itemize}
then $\mathit{proc\_state\_related}\: M' \:(\texttt{ADC\_Coq}\: (\mathit{arguments},\:st))$.
\end{theorem}

In the Coq formal model of ARM, transitions are terminating functions
returning a result of type \texttt{result}, 
as defined in section~\ref{sec:armmodel}.

The proof process is driven by the structure of the operation body. 
Step by step, we observe the memory changes on the Compcert-C side 
and the state changes on the Coq side, 
and we check whether the relation still holds between 
the current Compcert-C memory state and the Coq state.
To this effect, 
we apply theorems on load/store functions from Compcert~\cite{lerbla08}.
Proof by computation does not work because the types involved
are complex -- they embed logical information --
and many definitions are opaque.

In \texttt{ADC\_Coq}, conditional expressions and function calls for getting values 
have no side effect on the state.
On the Compcert-C side, declaring a local variable in a function has no impact on
the memory model of the processor. 
The state may only change when a function for setting values is called, 
like \texttt{set\_reg}, \texttt{copy\_StatusRegister}, 
or assignment of bits in register fields. 
Such calls will return a new memory state on the Compcert-C side
and a new \texttt{Ok} state on the Coq side.
We use small-step semantics for such steps.

Now we need some lemmas for these proof steps. 
Lemmas can be organized into four kinds.
We give an instance of each kind.

\begin{lemma}\label{l:1}
The conditional expression \texttt{S==1} has no effect on Compcert-C memory state:\\
if $G,E\vdash \texttt{condition}_C ? a1 : a2, M\overset{E0}{\Rightarrow}vres,M'$,\\
then M=M'.
\end{lemma}
Lemma \ref{l:1} is easy to prove by some inversions. 
All lemmas of this kind have been discharged.

\begin{lemma}\label{l:2}
The conditional expression \texttt{S==1} 
has the same result in the Compcert-C model as in the Coq model:\\
if $G,E\vdash \texttt{condition}_C ? a1 : a2, M\overset{E0}{\Rightarrow}vres,M'$:\\
- and if $is\_true~vres$, then $condition_{Coq} = true$\\
- and if $is\_false~vres$, then $condition_{Coq} = false$.
\end{lemma}
To prove lemma \ref{l:2}, we need to apply small-step semantics, 
to check the type of \texttt{S} and 
the value of the Boolean result \texttt{vres}.
Note that in Compcert-C, 
non-zero integer, non-zero floats and non-null pointer can
be interpreted as the Boolean value true,
which adds some complexity in the proof.

The proof is by case analysis according to the type of \texttt{vres}.
As the expression involves a parameter ($S$),
the projective relation about this parameter
between Compcert-C memory and the formal model of ARM is required.

All lemmas of this kind have been discharged.

%

\medskip
A lemma of the two next kinds is stated for each \simlight library function 
which changes the state,
e.g., \texttt{set\_reg}.
\begin{lemma}\label{l:3}
If $\mathit{proc\_state\_related} \:M~(\texttt{Ok}~st)$,\\ 
and if
$G,E\vdash \texttt{set\_reg}_c(proc,reg\_id,data), M\overset{E0}{\Rightarrow}vres,M'$,\\
then $\mathit{proc\_state\_related}~ M'~ (\texttt{set\_reg}_{Coq}~st)$.
\end{lemma}
For the moment, such lemmas are considered as axioms on the library.
In order to be properly stated, 
we need the Compcert-C ASTs of such library functions,
which are not automatically generated.
We have 6 lemmas/axioms of this kind for \texttt{ADC}.

\medskip
The next lemma is stated for
a given call to \texttt{set\_reg} in the body of the function 
\texttt{ADC\_Coq\_simlight}
and a parameter $P$ of \texttt{ADC\_Coq\_simlight} 
which is not used as an argument of  \texttt{set\_reg}.
\begin{lemma}\label{l:4}
After the call to \texttt{set\_reg},
the value of $P$ remains unchanged:\\
if 
$G,E\vdash \texttt{set\_reg}_c(proc,reg\_id,data), M\overset{E0}{\Rightarrow}vres,M'$,\\
then $P(M)=P(M')$.
\end{lemma}
Lemma 4 can be proved with the help of theorems of Compcert 
on ``load after store''. A typical proof step looks like:
\begin{quote}
If we store a value $v$ on block $b$ ($\texttt{store}(M1,\tau,b,\ofs,v)=M2$),
then 
the contents of block $b'$ remains unchanged
($\texttt{load}(\tau',M2,b',\ofs')=\texttt{load}(\tau',M1,b',\ofs')$)
for any type $\tau'$ and offset $\ofs'$,
which makes the accesses disjoint
($b'\neq b$ or $\ofs+|\tau|\leq \ofs'$ or $\ofs'+|\tau'|\leq \ofs$).
\end{quote}
As for lemmas \ref{l:3}, we need additional axioms on \simlight library functions.

\medskip
Our current result is that, with the help of these lemmas, 
we have a complete correctness proof for \texttt{ADC} (Theorem \ref{t:adc}).
Theorem \ref{t:adc} 
and all the lemmas are in the file \verb|correctness_ADC.v|$^{\ref{f:site}}$.

The whole proof structure of this theorem and 
all twenty lemmas of kinds \ref{l:1} and \ref{l:2} were completed within 2 weeks.
The 10 remaining lemmas, of kinds \ref{l:3} and~\ref{l:4}, should require a similar
effort. 
Here, we first need to generate Compcert-C ASTs for the 
relevant library functions using the C parser available in Compcert.


\section{Conclusion}
\label{sec:conclusion}

The trust we may have in our result depends on
the faithfulness of its statement with relation to 
the expected behavior of the simulation of \texttt{ADC} in \simlight.
It is mainly based on 
the manually written Coq and C library functions, 
the translators written in OCaml described in 
Section~\ref{sec:overallarchi}
(including the pretty-printer for Coq),
the final phase of the Compcert compiler,
and the formal definition of $\mathit{proc\_state\_related}$.

The current development is available online\footnote{%
\label{f:site}
\url{http://formes.asia/media/simsoc-cert/}}.
Figures on the size of our current development are given in
Table~\ref{tab:sizes}.

\begin{table}[t]
  \centering
  \begin{tabular}{|l|r@{~}|}
    \hline 
    Original ARM ref man (txt)                                                           & 49655 \\ 
    ARM Parsing to an OCaml AST                                                          & 1068 \\ 
    Generator (Simgen) for ARM and SH with OCaml and Coq pretty-printers                 & 10675 \\ 
    Generated C code for Simlight ARM operations                                         & 6681 \\ 
    General Coq libraries on ARM                                                         & 1569 \\ 
    Proof script on \texttt{ADC}                                                                  & 1461 \\ 
    \hline 
  \end{tabular}
  \smallskip
  \caption{Sizes (in number of lines)}
  \label{tab:sizes}
\end{table}

In the near future, we will extend the work done on \texttt{ADC} 
to all other operations. 
The first step will be to design relevant suitable tactics,
from our experience on \texttt{ADC}, 
in order to shorten a lot the current proof and make it easier
to handle and to generalize.
We are confident that the corresponding work on 
the remaining ARM operations will then be done much faster,
at least for arithmetical and Boolean operations.

Later on, we will consider similar proofs for the decoder
-- as for the body of operations, it is already automatically
extracted from the ARM reference manual.
Then a proven simulation loop (basically, repeat decoding
and running operations) will be within reach.

In another direction, we also reuse the methodology based on
automatic generation of simulation code and Coq specification
for other processors. 
The next one which is already considered is SH4. 
In fact, the same approach as the ARMv6 has been followed, and a
similar Coq representation can currently be generated from the SH4
manual. Moreover, as the SH pseudo-code is simpler than the ARM, we are
hence impatient to work on its equivalence proof.

\subsection*{Acknowledgement}
We are grateful to Vania Joloboff and Claude Helmstetter for
their many explanations on SimSoC.
We also wish to thank the anonymous reviewers for their
detailed comments and questions. 


\bibliographystyle{abbrv}
\bibliography{biblio}

\end{document}